\documentclass[pra,twocolumn]{revtex4}
\usepackage{amstext,amsmath,amssymb,amsfonts}
\usepackage{latexsym}
\usepackage{euscript}

\DeclareMathOperator{\tr}{tr}

\def\be{\begin{equation}}
\def\ee{\end{equation}}
\def\bes{\begin{eqnarray}}
\def\ees{\end{eqnarray}}

\newcommand{\SU}{\mathrm{SU}}

\def\6{\langle}
\def\9{\rangle}
\def\tr{{\rm tr}\,}
\def\half{\mbox{$1\over2$}}
\def\ha3{\mbox{$3\over2$}}
\newcommand{\1}{\mathbb{1}}

\def\1{\mbox{1\hskip-.25em l}}

\begin{document}
\title{From qubits to black holes: entropy, entanglement and all that}
\author{
 Daniel R. Terno\footnote{dterno@perimeterinstitute.ca} }
\affiliation{Perimeter Institute for Theoretical Physics, 31 Caroline St, Waterloo, Ontario, Canada N2L 2Y5}

\begin{abstract}

 Entropy plays a
 crucial role in characterization of information and entanglement,
 but it is not a scalar quantity and for many systems it is
 different for different relativistic observers. Loop quantum gravity
 predicts the Bekenstein-Hawking term for black hole entropy and
 logarithmic correction to it. The latter originates in the entanglement
 between the pieces of spin networks that
 describe black hole  horizon. Entanglement between gravity
 and matter may restore the unitarity in the black hole evaporation process.
 If the collapsing  matter is assumed to be initially in a pure
 state, then  entropy of the Hawking radiation is exactly the
 created entanglement between matter and gravity.

\end{abstract}
\maketitle
Quantum mechanics and relativity theory emerged at the beginning
of the twentieth century to provide answers to unexplained issues
in physics: the blackbody spectrum, the structure of atoms and
nuclei, the electrodynamics of moving bodies. Many years later,
information theory was developed  for analyzing the efficiency of
communication methods.  These seemingly disparate disciplines are
actually intimately linked. Information arguments of different
degrees of formality were made in the analysis of wave
propagation, thermodynamics and black hole physics. The advent of
quantum information theory \cite{asher,nc} and its extension to
the relativistic domain
\cite{pt04} made this link even more apparent.

 Quantum information theory is based on two central concepts.
  The ultimate quantum nature of the bearers of
information is   formalized in a notion of a
\textit{qubit}, a quantum two-level system that supplants a bit of
the classical information theory \cite{nc}. A quantitative
approach to
 to the exhibition of stronger-than-classical correlations
between the subsystems led to the \textit{entanglement} theory
\cite{asher}. For pure states the concept of entanglement is straightforward:
 any state that is not  a direct product  is
 an entangled state, like the singlet of two qubits
\be
|\Psi^-\9_{AB}=\frac{1}{\sqrt{2}}\left(|01\9-|10\9\right).\label{singlet}
\ee
Entropy is central for quantification of information and
entanglement. A pure state entanglement has a canonical measure---
the degree of entanglement--- which is the von Neumann entropy of
either of the reduced density matrices, e.g. $E(\Psi^-_{AB})\equiv
S(\tr_A\Psi^-_{AB})=1~\rm{ebit}$.  The mixed state entanglement
theory is much more complicated and is far from being fully
developed
\cite{nc}.

 While  it is
 a resource in quantum information processing demanding careful
establishing, preservation and manipulation, the entanglement is
ubiquitous in physics. A formal decomposition of the Minkowski
vacuum in  ``inside" and ``outside" degrees of freedom
\cite{bomb},
\be
|\Omega\9\simeq\sum_{ij}c_{ij}|\psi_i^A\9\otimes|\psi_j^B\9,
\label{decompose}
\ee
can be done explicitly in the case of Unruh effect, with the
Minkowski vacuum being a maximally entangled state across all the
Rindler modes,
\be
|\Omega\9\simeq\prod_i\sum_{n=0}^\infty
\exp(-n\pi\omega_i/a)|n_{iA}\9\otimes|n_{iB}\9.
\ee

Tools and concepts  of quantum information may be usefully applied
to the questions of relativistic physics \cite{qitapl1}, black
hole thermodynamics and quantum gravity \cite{qitapl2}. A
necessary preliminary step, however, is  to re-examine  its basic
notions  in
 the relativistic setting \cite{pt04}.

 Intuitively entropy is thought to be the system's property and to be
invariant (for example, under Lorentz transformations). A more
careful analysis shows that this is not necessarily true. For
example qubits are usually  realized as spins   of massive
particles or polarizations of photons. Hence a state of a physical
qubit is obtained by tracing out particle's momentum. However,
Lorentz transformations of discrete degrees of freedom typically
depend on the momenta, thus making qubit density matrix
non-covariant, spin entropy non-invariant and the entanglement
between spins of several particles observer-dependent
\cite{pt04}. Since the overall transformation
 is unitary the information is not lost  but its distribution
 between different degrees of freedom depends on the observer.

 Similarly, a reduced density matrix $\rho^A_{ij}=\sum_k c_{ik}c^*_{jk}$
 that corresponds to  tracing out the
``outside" degrees of freedom in Eq.~(\ref{decompose})  (that are
labelled by ``B") is not covariant under Lorentz transformations.
For $\rho$ to have  a definite transformation law, it  is
necessary for  a unitary transformation on the entire space,
$U(\Lambda)$ to be a direct product of unitaries that act on the
``inside" and ``outside" spaces, $U(\Lambda)=U_A(\Lambda)\otimes
U_B(\Lambda)$. The vacuum state spans a one-dimensional
irreducible representation space of the Poincare group. The direct
product structure is compatible with this property of $|\Omega\9$
only if both representations $U_A$ and $U_B$ are one-dimensional.
However, the ensuing product structure of the vacuum  is
incompatible with  long-range correlations in it, and, in
particular, with the violation of the Bell-type inequalities
\cite{pt04,qitapl1}. A similar argument applies to all other states. Hence, unlike the full state
that transforms unitarily, its local reduced density operator does
not have a definite Lorentz transformation law, and its entropy is
not necessarily invariant. A byproduct of the non-invariance of
entropy is that the (effective) number of degrees of freedom,
$N=e^{S_{\max}}$, is observer-dependent \cite{dt04}.

An isolated stationary black hole furnishes an example of the
invariant entropy.  Consider two observers  at the same spacetime
point outside the horizon. Their local coordinate frames are
connected by some Lorentz transformation $\Lambda$. The global
role of this  Lorentz transformation is
 to define  new
surfaces of simultaneity. These surfaces may intersect the event
horizon differently from those of the original observer. However,
the area theorem of Hawking
guarantees that all these intersections lead to the same horizon
area.  Since  black hole entropy is a function of the area its
invariance keeps the entropy invariant
\cite{dt04,bhcft}.

  For a
long time entanglement has been connected to the black hole
entropy calculations
\cite{pt04,bomb}.  Loop quantum gravity expresses the black hole entropy in terms of the
object  itself \cite{lqg}, providing both the Bekenstein-Hawking
term $S\propto A/4$ and the logarithmic correction to it. While
there is still a controversy on the precise value of the
logarithmic correction, we now argue that the correlations (and
entanglement) between the pieces of the spin network that
describes the horizon are responsible for it \cite{lit2}. This is
easy to see in the following simple model, while the conclusion is
valid in a generic scenario.

Considering the horizon as a closed surface, loop quantum gravity
describes it as made of patches of quantized area and describes
the interior of the black hole in terms of (a superposition of)
spin networks whose boundary puncturing the horizon defines the
patches of the horizon surface. For an external observer it does
not matter what is inside the black hole, and only the horizon
information is relevant.  $\SU(2)$ representations that are
indexed by $i=1,..,n$ and carry the  spins $j_i$ are attached to
the elementary patches one the horizon. The bulk geometry within
the horizon is fully described by a spin network, which can
possibly have support on a complicated graph within the horizon.
For the external observer the bulk spin network is fully
coarse-grained and  the state of the patches on the boundary
belongs to the tensor product $V^{j_1}\otimes..\otimes V^{j_n}$.
 The only
constraint on physical states is that they should be globally
gauge invariant, i.e. $\SU(2)$ invariant, such that the possible
horizon states are the intertwiners (invariant tensors) between
the representations $V^{j_i}$.

 Consider for simplicity the horizon that is  made of  $2n$ of spin 1/2
patches: a horizon state will live in the tensor product
$(V^{1/2})^{\otimes 2n}$. The admissible microstates are the
intertwiners between these $2n$ representations, so that the
horizon can be said to be made of $2n$ qubits and the microstates
are singlets of these $2n$ qubits. Finally, the area of the
horizon surface will by definition be $2n\times a_{1/2}$, where
$a_{1/2}$ is the area associated with an elementary patch.

Then the density matrix that is ascribed to the black hole horizon
is an equal mixture of all the intertwiners, and its entropy in
the limit of large $n$ is
\be
S(\rho)=-\tr\rho\log\rho\sim 2n\log2-\frac{3}{2}\log n.
\ee
The $\SU(2)$ invariance of the horizon states forces $\rho$ to be
entangled.  Up to date there is no universal criterion to decide
whether a mixed  state is entangled at all, as opposed to
containing only classical correlations, if it has a larger
dimension than a two-qubit state.  There are also practically no
analytic calculations of the entanglement measures. However, the
entanglement of the horizon state $\rho$ was calculated
\emph{exactly}
\cite{lit1}. For a splitting of the spin network into two equal
parts it is $E(\rho)\sim\half\log n$, and the entropy of each half
is $S(\rho_n)\sim n\log2$. The quantum mutual information
$I_\rho(A:B)$ is the difference between the entropies of the
subsystems and the entire system and measures  the total amount of
classical and quantum correlations. (For a pure entangled state
$\Psi_{AB}$  it is just $I_\Psi(A:B)\equiv 2E(\Psi_{AB})$. Because
of the presence of classical correlations in  mixed entangled
states it should be greater). Indeed, for the black hole horizon
$I_\rho(n:n)\sim 3E(\rho)$ \cite{lit2,lit1}, so the corrections to
the semiclassical area law are the expression of quantum
entanglement.

While it is not obvious that the unitarity \emph{must} persist in
the process of creation and evaporation of black holes,
consideration  of the matter alone  is not sufficient to
convincingly preserve it
\cite{pt04, unibh}. Entanglement between gravitational and matter degrees
of freedom may the way to restore unitarity.

We outline the simplified unitary scenario. Initially the
spacetime is approximately flat and the matter is in some state
$\rho$. We describe it by a state $\Phi$ that corresponds to a
classical nearly Minkowski metric. The evolution that ends in the
black hole evaporation is unitary and is schematically described
as
\be
\Xi=U(\Phi\otimes\rho )U^{\dag},
\ee
where $\Xi$ is the final \emph{entangled} state of matter and
gravity. Reduced density operators  give predictions for the
gravitational background and the matter distribution on it. The
evolution of matter is obtained by tracing out the gravitational
degrees of freedom and is a completely positive non-unitary map
\cite{nc}. If we assume that the initial states are pure, then the
entropy of a reduced density operator is exactly the degree of
entanglement between matter and gravity, $E(\Xi)$. Hence, the
increase in the entropy of matter is not an expression of
information loss, but  a measure of the created entanglement, i.e.
\emph{redistribution} of information.

\medskip

The author is grateful to Viqar Hussain and Oliver Winkler for
helpful discussions and to Etera Livine for encouragement,
collaboration and critical comments.


\begin{thebibliography}{99}


\bibitem{asher} Peres A.  (1993) {\it Quantum Theory: Concepts and Methods\/}
(Kluwer, Dordrecht, The Netherlands).
\bibitem{nc}  Nielsen, M. A.  and  Chuang, I. L. (2000) {\it
Quantum Computation and Quantum Information\/}  (Cambridge
University Press, New York).
\bibitem{pt04}  Peres, A.  and  Terno, D. R. (2004)  \emph{Rev. Mod. Phys.} {\bf 76},
93.

\bibitem{bomb} Bombelli, L. R., Koul, R.,  Lee, J.,  and R. Sorkin (1986) \emph{Phys. Rev.
 D} {\bf34}, 373.

\bibitem{qitapl1} Summers, S. J. and Werner, R. (1985) \emph{Phys. Lett. A} {\bf
110}, 257;  Terno, D. R. (2003) \emph{Phys. Rev. A} {\bf 67},
014102.
\bibitem{qitapl2} Bekenstein, J. D. (1993) \emph{Phys. Rev. Lett} {\bf70}, 3680;
Terno, D. R. (2002) in {\it Quantum Theory: Reconsideration of
Foundations\/}, edited by A. Khrennikov (V\"axj\"o University,
V\"axj\"o, Sweden) p.~397; Hawkins, E.,  Markopoulou, F., Sahlman,
H. (2003) \emph{Class. Quant. Grav.} {\bf 20}, 3839;  Dreyer, O.,
Markopoulou, F., and Smolin, L. (2004) hep-th/0409056.

\bibitem{dt04}  Terno, D. R. (2004) \emph{Phys. Rev. Lett.} {\bf 93},
051303.
\bibitem{bhcft} Fiola T. M. , Preskill, J.,  Strominger, A., and Trivedi, S. P.  (1994)
\emph{Phys. Rev. D} {\bf 59}, 3987.

\bibitem {lqg}  Rovelli, C. (2004) {\em Loop Quantum Gravity}, (Cambridge University
Press, Cambridge);  Thiemann, T. (2003) {\em Lectures on Loop
Quantum Gravity}, Lect. Notes Phys. {\bf 631},  41;   Ashtekar, A.
and Lewandowski, J. (2004)  Class. Quant. Grav. {\bf 21}, R53.

\bibitem{lit2}  Livine,  E. R.  and Terno, D. R. (2005) preprint.
\bibitem{lit1} Livine,  E. R.  and Terno, D. R. (2005) quant-ph/0502043.
\bibitem{unibh}  Gottesman, D. and Preskill, J. (2004) \emph{JHEP} {\bf
0403}, 026; Gambini, R., Porto, P. R., and Pullin, J. (2004)
\emph{Phys. Rev. Lett.} {\bf 93}, 240401.

\end{thebibliography}
\end{document}